\documentclass[
    ,final            
  ]
  {aipproc}

\layoutstyle{6x9}

\def\apj{ApJ}

\def\mnras{MNRAS}

\def\aj{AJ}

\newcommand{\eg}{{\it e.g.}}

\newcommand{\hi}{{H{\sc i}}}
\newcommand{\rband}{{\em r}-band}

\newcommand{\zband}{{\em z}-band}

\newcommand{\x}{$\times$}
\newcommand{\about}{$\sim$}
\newcommand{\Msun}{M$_\odot$}

\newcommand{\Mhi}{$M_{\rm HI}$}
\newcommand{\Mst}{$M_\star$}
\newcommand{\must}{$\mu_\star$}
\newcommand{\nuvr}{NUV$-r$}
\newcommand{\Rinz}{$R_{50,z}$}

\begin{document}

\title{HI and Star Formation Properties of Massive Galaxies: 
First Results from the GALEX Arecibo SDSS Survey}



\author{Barbara Catinella}{
  address={Max-Planck Institut f\"{u}r Astrophysik, D-85741 Garching, Germany}
}

\author{David Schiminovich}{
  address={Department of Astronomy, Columbia University, New York, NY 10027, USA}
}

\author{Guinevere Kauffmann}{
  address={Max-Planck Institut f\"{u}r Astrophysik, D-85741 Garching, Germany}
}

\begin{abstract}
The GALEX Arecibo SDSS Survey (GASS) is an ambitious program
designed to investigate the cold gas properties of massive galaxies, a
challenging population for \hi\ studies. 
Using the Arecibo radio telescope, GASS is gathering high-quality \hi-line spectra
for an unbiased sample of \about 1000 galaxies with stellar masses
greater than $10^{10}$ \Msun\ and redshifts $0.025 < z < 0.05$,
uniformly selected from the SDSS spectroscopic and GALEX imaging surveys. 
The galaxies are observed until detected or until a low gas mass
fraction limit (1.5$-$5\%) is reached. 
We present initial results based on the first Data
Release, which consists of \about 20\% of the final GASS sample.
We use this data set to explore the main scaling relations of
\hi\ gas fraction with galaxy structure and \nuvr\ colour, and show our 
best fit plane describing the relation between gas fraction, stellar mass 
surface density and \nuvr\ colour. Interesting outliers from this plane 
include gas-rich red sequence galaxies that may be in the process of regrowing their
disks, as well as blue, but gas-poor spirals.
\end{abstract}

\maketitle


\section{Introduction}

While the distinction between red, old ellipticals
and blue, star-forming spirals has been known for a long
time, recent work based on the Sloan Digital Sky Survey (SDSS, \cite{sdss}) 
has shown that galaxies appear 
to divide into two distinct ``families'' at a stellar mass 
\Mst \about 3 \x $10^{10}$ \Msun\ \cite{str01,kau03,bal04}.
Lower mass galaxies typically have young stellar populations, low
surface mass densities and the low concentrations characteristic of
disks. On the other hand, galaxies with old stellar populations, high
surface mass densities and the high concentrations typical of bulges
tend to have higher mass.
It is clearly important to understand why there should be a
characteristic mass scale where galaxies transition 
from young to old. And, in order to understand how such transition
takes place, it is critical to study the cold \hi\ gas, which is the 
source of the material that will eventually form stars. 
\hi\ studies of transition objects require large and uniform samples 
spanning a wide range in gas fraction, stellar mass and other
galaxy properties (\eg, structural parameters and star formation).
Although blind surveys offer the required uniformity, \hi\ studies of
transition galaxies are currently not possible because the depths
reached by existing wide-area blind \hi\ surveys are very shallow
compared to surveys such as the SDSS.
The GALEX Arecibo SDSS Survey (GASS) is a new \hi\ survey specifically
designed to obtain \hi\ measurements of \about 1000 massive
galaxies in the local universe, selected only by redshift and 
stellar mass. The first Data Release (DR1) and initial results are presented in \cite{gass1}.
GASS is assembling the first statistically significant
sample of massive galaxies with homogeneously measured stellar masses,
star formation rates and gas properties. This unique data set will
allow us to investigate if and how the cold gas responds to a variety
of different physical conditions in the galaxy, thus yielding insights
on the physical processes that regulate gas accretion and its
conversion into stars in massive systems.

\section{Survey Design and Sample Selection}

The GASS targets are located within the intersection of the
footprints of the SDSS primary spectroscopic survey, the projected
GALEX Medium Imaging Survey and the Arecibo Legacy Fast ALFA (ALFALFA,
\cite{alfalfa}) \hi\ survey. Existing ALFALFA coverage
increases our survey efficiency by allowing us to remove from the GASS
target list any objects already detected by ALFALFA (\about 20\% of
the GASS sample). As already mentioned, the targets are selected only
by redshift ($0.025 < z < 0.05$) and stellar mass (\Mst $>$ $10^{10}$ \Msun). 
Our selected stellar mass range straddles the ``transition mass'' 
(\Mst \about $3 \times 10^{10}$ \Msun) above which galaxies
show a marked decrease in their present to past-averaged star formation rates.
The GASS targets are observed until detected or until a low gas mass fraction 
limit is reached. Practically, we have set a gas mass fraction limit of 
$M_{\rm HI}/M_\star > 0.015$ for galaxies with \Mst $> 10^{10.5}$ \Msun, and a
constant gas mass limit \Mhi $=10^{8.7}$ \Msun\ for galaxies with
smaller stellar masses. This corresponds to a gas fraction limit $0.015-0.05$ 
for the whole sample. This allows us to detect galaxies with gas
fractions significantly below those of the \hi-rich ALFALFA
detections at the same redshifts, and find early-type transition
galaxies harboring significant reservoirs of gas. 

Since the ALFALFA and GALEX surveys are on-going, we have defined a
GASS {\em parent sample} (12006 galaxies), based on SDSS DR6 and the maximal ALFALFA
footprint, from which the targets for Arecibo observations are
extracted. The final GASS sample will include \about 1000 galaxies,
chosen by randomly selecting a subset which balances the distribution
across stellar mass and which maximizes existing GALEX exposure time.

\section{Gas Fraction Scaling Relations}

In Figure~1, we show how the average \hi\ mass fraction of massive
galaxies varies as a function of stellar mass, stellar mass surface
density (defined as  $\mu_\star = M_\star/(2 \pi R_{50,z}^2)$, where
\Rinz\ is the radius containing 50\% of the Petrosian flux in \zband,
in kpc units), concentration index (defined as $R_{90}/R_{50}$, where
$R_{90}$ and $R_{50}$ are the radii enclosing 90\% and 50\% of the
\rband\ Petrosian flux, respectively) and observed \nuvr\ colour
(corrected for Galactic extinction only).
Large circles and triangles represent average and median gas fractions
in a given bin (computed after correcting the sample for the fact that
we do not re-observe objects already detected by
ALFALFA; see \cite{gass1} for details). We included the non-detections
with their \hi\ masses set to either zero (filled circles) or to their
upper limits (empty circles and triangles).
As can be seen, the answer is insensitive to the way we treat
the galaxies without \hi\ detections, except for the very most massive,
dense and red galaxies. For comparison, we also show galaxies in the
GASS parent sample detected by ALFALFA (dots). It is clear that the
shallower, blind \hi\ survey is biased to significantly higher gas
fractions compared to our estimates of the global average. 
As these plots show, the gas content of massive galaxies decreases
with increasing \Mst, \must, concentration index, and observed \nuvr\ colour.
The strongest correlations are with \must\ and \nuvr.
We also notice that the difference between the mean
and median values of \Mhi/\Mst\ is smallest when it is plotted as a function
of \must\ and \nuvr. This is because these two properties 
yield relatively tight correlations without significant tails to low values
of gas mass fraction. \\

\begin{figure}
  \includegraphics[height=.46\textheight]{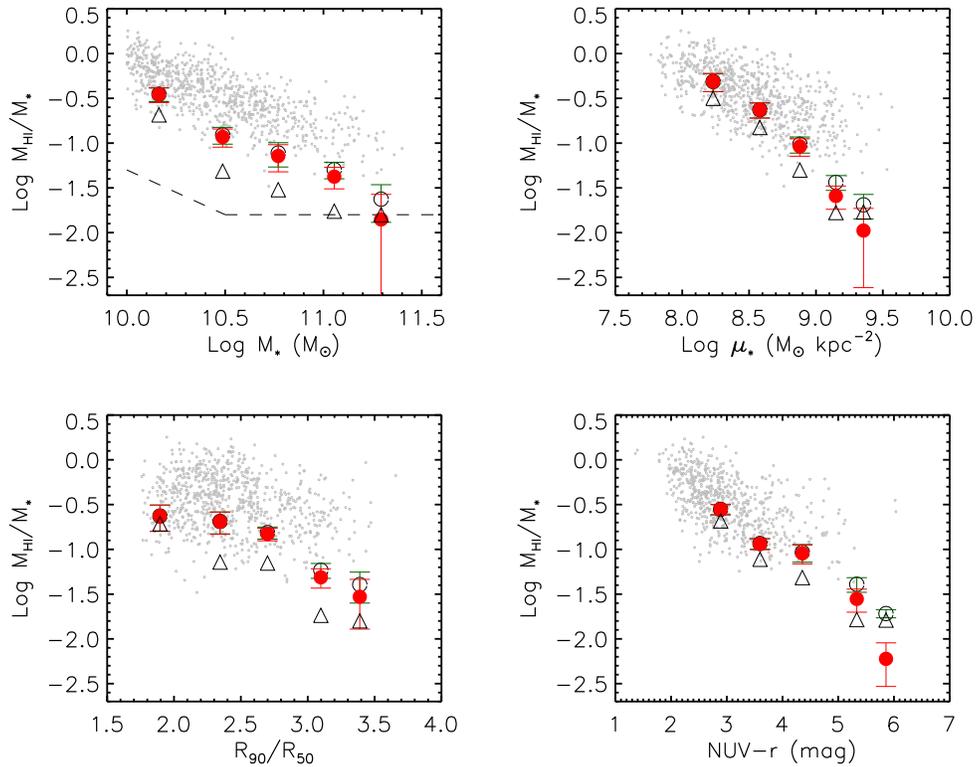}
  \caption{Average trends of \hi\ mass fraction as a function of stellar
mass, stellar mass surface density, concentration index and observed
\nuvr\ colour. In each panel, large circles indicate average gas
fractions. These were computed including the
non-detections, whose \hi\ mass was set to either its upper limit
(empty) or to zero (filled). Triangles are medians.
Error bars are from bootstrapping. Galaxies in the
GASS parent sample detected by ALFALFA are
plotted as dots. The dashed line in the first panel
shows the \hi\ detection limit of the GASS survey.}
\end{figure}

One of the key goals of the GASS survey is to identify and
quantify the incidence of {\em transition} objects, which might be
moving between the blue, star-forming cloud and the red sequence of
passively-evolving galaxies. Depending on their path to or from the
red sequence, these objects should show signs of recent quenching
of star formation or accretion of gas, respectively. 
In order to establish what is the {\em normal} gas content of a 
galaxy of given mass, structural properties and star formation rate, 
we have fit a plane to the 2-dimensional relation between \hi\ mass fraction,
stellar surface mass density, and \nuvr\ colour. 
Objects that deviate strongly from the average behavior of the sample are
the best candidates for galaxies that might be transitioning between
the blue and the red sequences.
The gas fractions obtained from our best fit relation are compared
with measured ones in Figure~2. ALFALFA galaxies and non-detections 
were not used in the fit and are shown for comparison only.
Galaxies which are anomalously gas-rich given
their colours and densities scatter above the mean
relation, while those that are gas-poor scatter below. This is clearly
demonstrated by the \hi-rich ALFALFA galaxies, which are preferentially
found above the line. Marked on the diagram is GASS 3505 (filled star),
a galaxy that has optical morphology and colours characteristic of a
normal elliptical, but a 50\% \hi\ mass fraction. Also interesting
are the galaxies with low \hi\ mass fractions, but
that are still forming stars. These galaxies are found near the bottom
the plot, but shifted to the right, as exemplified by GASS 7050
(empty star), a gas-poor disk galaxy that was not detected in \hi.
These may be systems where the
\hi\ gas has recently been stripped by tidal interactions or by
ram-pressure exerted by intergalactic gas, or where other feedback
processes have expelled the gas. 
In future work, we plan to investigate these different classes of
transition galaxy in more detail.



\begin{figure}
  \includegraphics[height=.3\textheight]{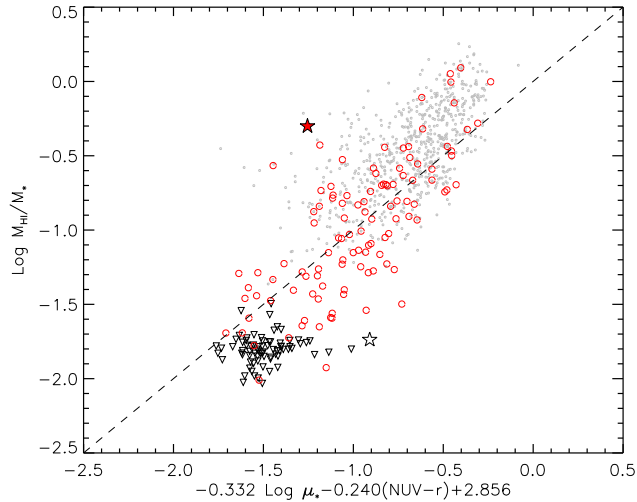}
  \caption{The best fit ``plane'' describing the relation between \hi\ mass fraction, stellar mass
surface density and observed \nuvr\ colour. Circles represent
\hi\ detections, upside-down triangles are non-detections, dots are
galaxies in the GASS parent sample detected by ALFALFA. The 1:1 relation is
indicated by a dashed line. Stars are discussed in the text.}
\end{figure}




\begin{thebibliography}{6}

\bibitem[\protect\citeauthoryear{York et~al.}{2000}]{sdss}
D.~G. {York}, J.~{Adelman}, J.~E. {Anderson}, Jr., et~al., \emph{\aj} \textbf{120}, 1579--1587 (2000).

\bibitem[\protect\citeauthoryear{Strateva et~al.}{2001}]{str01}
I.~{Strateva}, {\v Z}.~{Ivezi{\'c}}, G.~R. {Knapp}, et~al., \emph{\aj} \textbf{122}, 1861--1874 (2001).

\bibitem[\protect\citeauthoryear{Kauffmann et~al.}{2003}]{kau03}
G.~{Kauffmann}, T.~M. {Heckman}, C.~{Tremonti}, et~al., \emph{\mnras} \textbf{346}, 1055--1077 (2003).

\bibitem[\protect\citeauthoryear{Baldry et~al.}{2004}]{bal04}
I.~K. {Baldry}, K.~{Glazebrook}, J.~{Brinkmann}, et~al., \emph{\apj} \textbf{600}, 681--694 (2004).

\bibitem[\protect\citeauthoryear{Catinella et~al.}{2009}]{gass1}
B.~Catinella, D.~Schiminovich, G.~Kauffmann et~al., \emph{MNRAS} in press, arXiv:0912.1610 (2009).

\bibitem[\protect\citeauthoryear{Giovanelli et~al.}{2005}]{alfalfa}
R.~{Giovanelli}, M.~P. {Haynes}, B.~R. {Kent}, et~al., \emph{\aj} \textbf{130}, 2598--2612 (2005).


\end{thebibliography}
\end{document}